\documentclass[aps,10pt,prl,twocolumn,amsmath,amssymb]{revtex4-1}

\usepackage{graphicx}
\usepackage{dcolumn}   
\usepackage{bm}        
\usepackage{color}
 \newcommand{\change}[1]{{#1}}

\begin{document}

\title{Attosecond Photoscopy of Plasmonic Excitations}

\author{Mattia Lupetti$^a$, Julia Hengster$^b$, Thorsten Uphues$^b$, and Armin
  Scrinzi$^a$} \email{mattia.lupetti@physik.uni-muenchen.de}
\email{armin.scrinzi@lmu.de} \affiliation{$^a$Physics Department,
  Ludwig Maximilians Universit\"at, D-80333 Munich, Germany}
\affiliation{$^b$Center for Free-Electron Laser Science, Universit\"at
  Hamburg, D-22761 Hamburg, Germany}


\date{\today}

\begin{abstract}
  \noindent We propose an experimental arrangement to image, with
  attosecond resolution, transient surface plasmonic excitations. 
  The required modifications to state-of-the-art setups used for
  attosecond streaking experiments from solid surfaces only involve
  available technology. \change{Buildup and life times of surface plasmon
  polaritons can be extracted and local modulations of the exciting optical 
  pulse can be diagnosed {\it in situ}.}
\end{abstract}

\maketitle


\change{Surface plasmons} are collective excitations of
electrons that propagate along a metal-dielectric interface.
Recently, plasmonics has gathered interest for the development of
ultra-fast all-optical circuitry \cite{Ozbay2006}, since it can
combine the high operational speed of photonics (PHz scale) with the
miniaturization provided by electronics (nm scale). For this purpose,
it is important to understand the buildup dynamics and lifetime of the
collective electronic excitation. Although the plasmon lifetime can
be inferred from the plasmonic resonance width (of the transmission
spectrum, see for instance \cite{ropers2005prl}), plasmon buildup is 
a process that cannot be addressed in terms of frequency analysis. 

In the present work, we propose an experimental setup to image the transient dynamics of
a plasmonic mode, which can be realized as a modification of the so-called
``attosecond streak camera'' \cite{Kienberger2004}, which has
already been successfully applied to solid surfaces. The attosecond
streak camera is a two-color pump-probe scheme, where a weak XUV
attosecond pulse ionizes electrons from the solid, and a collinear,
few-cycle ($\sim 5\,fs$ FWHM) NIR pulse serves as the probe, which
accelerates the XUV photo-electrons after their escape from the
solid. With this technique it was possible to
resolve solid-state physics phenomena with resolution of a few
attoseconds ($1\, \text{as} = 10^{-18}$ s) \cite{Cavalieri2007}.

We benchmark our setup concept against the buildup of Surface 
Plasmon Polaritons (SPPs) excited by a NIR pulse on a grating surface. A
time-delayed XUV pulse probes the SPPs during their
evolution by detecting the effect of their field on XUV
photo\-emission. 
In principle, pump and probe beams can be spatially separated, 
allowing to probe different surface regions. 
Thus, differently from atomic and surface streaking employed
so far, the setup provides spatio-temporal information. To
distinguish it from standard attosecond streaking experiments, 
we name our setup ``attosecond photoscopy''.


A well established method for producing isolated attosecond pulses 
is the generation of high harmonic radiation (HHG) in noble gases \cite{Hentschel2001, Cavalieri2007, Corkum1993, Agostini2004}. 
An intense few cycle NIR laser pulse is focused
into a noble gas target and generates high harmonics of the
fundamental radiation. The XUV radiation co-propagates with the
driving laser pulse.  Both pulses are focused onto a sample with a
delayable two part mirror composed of an XUV multilayer mirror in the
inner part and a broadband NIR mirror in the outer part. The
multilayer mirror is designed as a high pass filter for the harmonics, 
which results in an isolated attosecond
pulse. The pulse can be timed relative to the NIR with a precision of
$\lesssim 10\,as$.

\begin{figure}[ht!]
  \includegraphics[width=\linewidth]{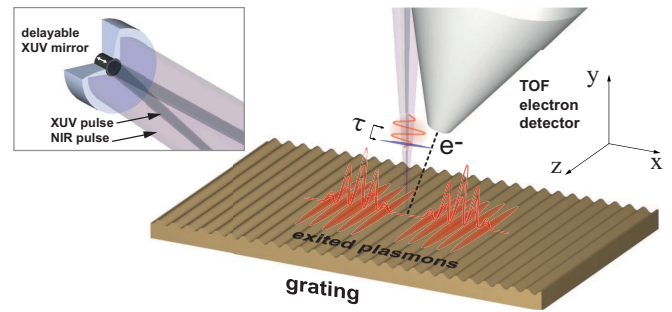}
  \caption{Experimental setup of an attosecond photoscopy
    experiment. The XUV attosecond pulse liberates electrons in presence of the
    plasmonic field, which is excited by a short NIR pulse. Control of
    NIR-XUV time delay $\tau$ allows observation of the plasmon transient
    dynamics.}
  \label{fig:setup}
\end{figure}

Figure \ref{fig:setup} illustrates the setup discussed here.
The NIR and XUV beams propagate in $y$-direction, at normal incidence 
onto the plane of the grating.
Polarizations are in $x$-direction, perpendicular to the grooves. Using this
arrangement, two counter-propagating plasmons are excited in the focus
of the NIR pulse on the grating structure. 
\change{A band gap at the zero crossing separates two plasmon branches \cite{barnes1996prb}. 
An optical pulse at normal incidence usually couples to only one of
the branches, called the bright mode, but at tight focussing
with about 5$^\circ$ angular dispersion also the second, ``dark'' mode is excited.}

XUV photo-electrons are measured at perpendicular direction to the
surface. As in \cite{Cavalieri2007}, the final electron momenta are
recorded as a function of the delay between the NIR and XUV beams.
The electron \emph{spectrogram} retrieved is a convolution of photoemission 
with acceleration in the plasmonic field at the location 
and time of the initial electron release.

Depending on the time delay between the NIR pulse and the probing
attosecond pulse, the XUV generated photoelectrons experience a
different plasmonic field amplitude and phase, leading to a
modulation of the kinetic energy distribution by the emerging plasmonic field.

\change{ The energy gap between dark and bright modes manifests itself 
in the spectrogram as a ``transition'' from the bright $\omega_b$ 
to the dark $\omega_d$ mode frequencies, 
which is measurable in our setup because of the attosecond resolution.}

Below we analyze the photoscopic spectrogram using a basic analytical
model as well as numerical solutions of the SPP propagation together
with a Monte Carlo simulation of the electron streaking process. We
will demonstrate that from the spectrograms one can recover the  
plasmonic field at the surface. The detailed analysis and
interpretation will be discussed in the following.


Standard streaking experiments are based on electron sources that can be
considered point-like with respect to the laser wavelength, such as 
atoms or molecules. For this reason the dipole
approximation can be used: $\mathbf{A}(\mathbf{r},t)\simeq
\mathbf{A}(t)$. After emission, the electron
canonical momentum is conserved: $ \mathbf{P}(t) = \mathbf{P}_i$,
which translates into $\mathbf{p}(t) + \frac{e}{c}\mathbf{A}(t) =
\mathbf{p}_i + \frac{e}{c}\mathbf{A}(t_i)$, where $e$ denotes the electron charge 
and $|\mathbf{p}_i| = \sqrt{2m(E_{xuv}-W_f)}$ is the initial momentum of the electron
released at time $t_i$ from a material with work function $W_f$.
Assuming that $A(t\rightarrow\infty) = 0$, the final momentum recorded
by the spectrometer is
\begin{equation}\label{eq:fin-mom}
  \mathbf{p}_f = \mathbf{p}_i + \mathbf{a}(t_i),
\end{equation}
where we defined $\mathbf{a}:= \frac{e}{c}\mathbf{A}$.

The spectral width of the XUV attosecond pulse is reflected in a
momentum-broadening of the initial electron distribution $n_e =
n_e(\mathbf{p}_i, t_i)$. For simplicity we assume Gaussian
distributions centered around
momentum $\mathbf{p}_0$ and time $t_0$, respectively, where $t_0$
denotes the time of peak XUV intensity on target.  With Eq.~(\ref{eq:fin-mom}) for
the initial electron momentum, the time-integrated
final momentum is
\begin{equation}\label{eq:mom-dis}
  \sigma(\mathbf{p}_f)  = \int_{-\infty}^{\infty}\!\!\! dt_i \, n_e(\mathbf{p}_f - \mathbf{a}(t_i), t_i).
\end{equation}
The spectrogram for a series of delays $\tau$ becomes
\begin{equation}\label{eq:str-spe}
  \sigma(\mathbf{p}_f, \tau) = \int_{-\infty}^{\infty} dt_i \, n_e(\mathbf{p}_f - \mathbf{a}(t_i), t_i-\tau).
\end{equation}
From this, the NIR pulse can be reconstructed 
by analyzing the average momentum of the streaking
spectrogram

When applying the method to plasmonic excitations we have to
consider that the SPP, acting as the streaking field, is spatially inhomogeneous
and propagates on a surface. Previous
work on streaking on nanoparticles \cite{Suessmann2011prb} 
clearly shows that spatial inhomogeneity of the streaking field leads to a smearing of 
the streaking trace obtained in a traditional setup.  
Thus, we need to include the position dependence into our initial electron 
distribution:  
$n_e(\mathbf{p}_i, t_i)\rightarrow n_e(\mathbf{r}_i, \mathbf{p}_i,t_i)$. 
The final momentum of the electrons accelerated in the plasmon
field is then
\begin{equation}\label{eq:fin-mom-ext}
  \mathbf{p}_f = \mathbf{p}_i - e \int_{-\infty}^{\infty}\mathbf{E}(\mathbf{r}(t'), t')\, dt'.
\end{equation}
For a typical XUV photon energy of $80$ eV, the average initial speed
of a photoelectron is $v_i = 5$ nm/fs. If the NIR pulse is $4$
fs short, it will give rise to a plasmonic field of a duration
of few tens of femtoseconds.  During this time, the electrons
move by $\lesssim 100$ nm.  The additional drift imparted by the
plasmonic field is small compared to the initial velocity. As the
plasmon evanescent field extends to about NIR wavelength ($800$ nm)
beyond the surface, we can write $\mathbf{r}(t')\simeq
\mathbf{r}_i$ in Eq.~(\ref{eq:fin-mom-ext}). With this approximation,
one obtains a position corrected analog of Eq.(\ref{eq:fin-mom}):
\begin{equation}\label{eq:fin-mom-ext2}
  \mathbf{p}_f = \mathbf{p}_i - \mathbf{a}(\mathbf{r}_i, t_i)
\end{equation}
Since the photoelectron detector does not resolve the emission
positions $\mathbf{r}_i$, the photoscopic spectrogram is the integral
over time {\em and the area covered by the XUV pulse}
\begin{equation}\label{eq:pho-spe}
  \sigma(\mathbf{p}_f, \tau) = \int_{\mathbb{R}^3}d^3r_i
  \int_{-\infty}^{\infty} dt_i \, n_e(\mathbf{r}_i, \mathbf{p}_f - \mathbf{a}(\mathbf{r}_i, t_i), t_i-\tau).
\end{equation}
\change{The space-averaged momentum is independent of the time-delay, 
as the integral of a propagating pulse is negligible (exactly zero in free space). 
Thus} for extracting time information from the photoscopic spectrogram, we
use the delay-dependent momentum variance
\begin{equation}\label{eq:ps-mom2}
  S(\tau) = \frac{\int d\mathbf{p}_f \, |\mathbf{p}_f|^2
    \,\sigma(\mathbf{p}_f, \tau)}{\int d\mathbf{p}_f \,
    \sigma(\mathbf{p}_f, \tau)} - |\langle \mathbf{p}_f
  \rangle|^2.
\end{equation}
As the XUV pulse duration is short compared to the NIR period, 
we treat  photoemission as instantaneous. The distribution of the 
photoelectron yield along the surface is proportional to the XUV 
intensity profile. Furthermore, we neglect 
any transport effect in the solid and consider only the photoelectrons 
coming from the first few layers of material, as reported in
\cite{Neppl2012b}. With these conditions one finds
\begin{equation}\nonumber
  n_e(\mathbf{r}_i, \mathbf{p}_i, t_i-\tau)\simeq g_{\text{x}}(x_i)n_e(\mathbf{p}_i)\delta(y_i - y_s)\delta(t_i-\tau-t_0),
\end{equation}
where $y_s$ is the grating vertical position (we neglect any groove
depth effect) and $g_{\text{x}}$ is a Gaussian function of width
$w_\text{x}$, i.e. the XUV attosecond pulse focal spot.

As for the angular dependence of the photoemission we first restrict
our discussion to the two extreme cases of 1) unidirectional emission with
all initial momenta orthogonal to the grating plane, and 2)
isotropic emission.  
For either distribution, the reconstructed times closely
reproduce the actual dynamics.  In reality, the XUV photoelectron distribution 
will be between these extreme cases and should be
determined in a measurement without NIR field.

Unidirectional initial distributions can be written as
$n_e(\mathbf{p}_i) = n_e(p_i\, \hat{\mathbf{n}}_s)$, where $p_i =
|\mathbf{p}_i|$ and $\hat{\mathbf{n}}_s$ is the direction orthogonal
to the grating plane.  Eq.~(\ref{eq:pho-spe}) now
becomes
\begin{equation}\nonumber
  \sigma(p_f, \tau) = \int_{-\infty}^{\infty} dx_i\, g_\text{x}(x_i) n_e\left(p_f 
- \hat{\mathbf{n}}_s\cdot\mathbf{a}(x_i, t_0-\tau)\right),
\end{equation}
where $\hat{\mathbf{n}}_s$ denotes the surface normal.
Near the surface, in the region that is probed by the electrons, the
plasmonic field is predominantly perpendicular to the surface. Therefore, we
can approximate $\hat{\mathbf{n}}_s\cdot\mathbf{a} = \mathsf{a}_y
\simeq \mathsf{a}_{\text{spp}}$.
Computing the variance Eq.~(\ref{eq:ps-mom2}) for a Gaussian
distribution of the initial electron momenta, we obtain

\begin{align}
  S(\tau) &= \Delta p^2 + \int_{-\infty}^{\infty} dx_i\,
  g_\text{x}(x_i)\mathsf{a}_\text{spp}^2(x_i,
  t_0-\tau).\label{eq:ps-mom2-uni}
\end{align}
For isotropic XUV photo-electron emission, the initial distribution
can be written as: $n_e(\mathbf{p}_i) = \frac{1}{\pi} n_e(p_i)= \frac{1}{\pi}n_e(|\mathbf{p}_f - \mathbf{a}|)$, where
we employed $ p_i = |\mathbf{p}_i|$. We use
$|\mathbf{a}|\ll|\mathbf{p}_{f}|$ to approximate $|\mathbf{p}_f -
\mathbf{a}| \simeq p_{f} - \,\mathbf{a}\cdot\hat{\theta}$, where
$\theta$ is the angle between the final momentum and the surface
normal.  The spectrogram then reads
\begin{equation}\label{eq:pho-spe-iso}
  \sigma(p_f, \tau) = \frac{1}{\pi}\int_{-\infty}^{\infty} dx_i\, 
g_\text{x}(x_i)\, n_e(p_f - \mathbf{a}\cdot\hat{\theta}).
\end{equation}
A straightforward calculation for the angular integrations
leads to the expression of the variance
\begin{align}
  S(\tau) &= \Delta p^2 + \frac{1}{\pi}\int_{-\infty}^{\infty} dx_i\,
  g_\text{x}(x_i)|\mathbf{a}(x_i, \tau)|^2.
\label{eq:ps-mom2-iso}
\end{align}
In either case, by Eqs.~(\ref{eq:ps-mom2-uni}) and (\ref{eq:ps-mom2-iso}), 
measuring the variance of the photo-emission spectrogram provides direct access to
the space-averaged vector potential $\mathbf{a}^2$ at the surface in the direction of photo-detection.
The surface vector potential $|\mathbf{a}|^2 = \mathsf{a}_x^2 + \mathsf{a}_\text{spp}^2$
also includes $a_x$, the NIR field at the grating surface. 
Modifications of the surface field compared to the incident beam 
can be measured {\it in situ} (see below). 

Simulations of the plasmonic field were performed with the
finite-difference time-domain (FDTD) method \cite{Taflove1995}, using
a freely available software package \cite{Oskooi2010}. 
Material properties were included through the appropriate model of gold dielectric 
function \cite{Rakic98ao}.
We assume a Gaussian 4 fs FWHM pulse at a central wave length of 800 nm.
The grating parameters are optimized for maximal absorption from the NIR pulse,
assuming a gold surface. Beam waists of NIR and XUV were 5 and 10 $\mu$m, respectively.

The XUV photoemission process is approximated as a 
sudden ejection of electrons from the surface boundary, with the appropriate 
unidirectional and isotropic initial
momentum distribution, respectively. The electron trajectories and final
momenta are computed by solving the Lorentz equation for each
photoelectron in the previously simulated electromagnetic field. 

\begin{figure}[ht!]
  \centering
  \includegraphics[width=1.0\linewidth]{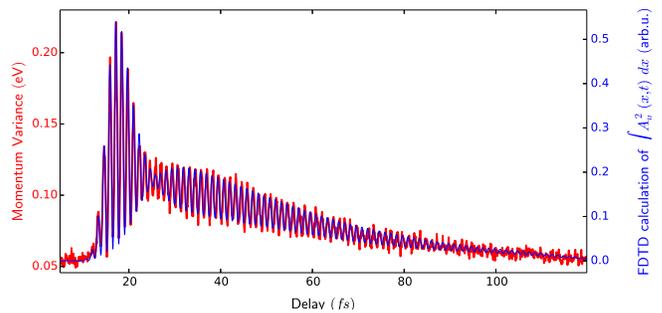}
  \vspace{-0.7cm}
  \caption{Comparison between variance of photoscopic spectrogram in the 
  "filtered isotropic" case (red) and $\int |\mathsf{a}_y|^2 dx$ computed in 
  the FDTD (blue). The offset of the filtered isotropic case is due to the 
  XUV pulse energy width.}
  \label{fig:comp_iso_sim}
\end{figure}

The spectrogram variance obtained by Monte Carlo simulation is compared in Fig. 
\ref{fig:comp_iso_sim} with the space integral of the squared vector potential 
along $y$ from the FDTD simulation. 
We assume isotropic initial momentum distribution and a TOF detector of 5$^\circ$
acceptance centered around the perpendicular direction.

Note that the variance directly images the integral of the surface plasmonic field 
squared without further assumptions or input from theory.
The agreement is robust w.r.t. to the angular distribution of photo-electron momenta: 
one obtains analogous results for unidirectional emission.

\begin{figure}[ht!]
  \centering
  \includegraphics[width=1.05\linewidth]{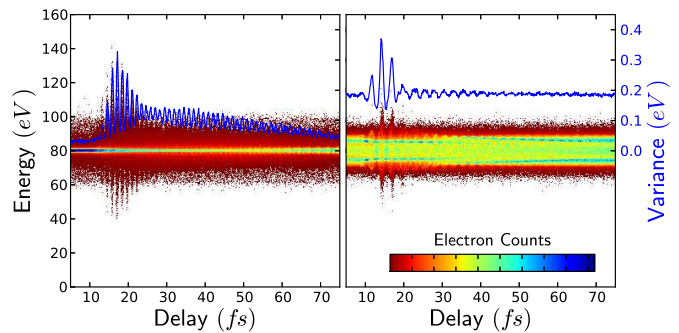}
  \vspace{-0.7cm}
  \caption{Photoscopic spectrograms at perpendicular (left) and grazing (right)
    electron emission. The measurements retrieve plasmonic and NIR field, respectively. 
    Solid lines are the momentum variances. }
  \label{fig:pho-spe-iso}
\end{figure}
The detailed image of the fields  provides for an {\it in situ} diagnosis both, of the 
plasmon field and exciting NIR source, including possible distortions due to the NIR 
reflection on the grating.
In Figure~\ref{fig:pho-spe-iso} spectrograms observed in the perpendicular and grazing
direction are shown, which reflect the two contributions.


From the plasmonic (perpendicular) component, we extract buildup and
life-times, as well as contributions of the bright and dark
modes to the spectrograms.
We parametrize the field as follows: we assume plasmonic fields with a Gaussian envelope
$\mathsf{a}_{\text{spp}} = \exp[i\varphi]\exp[-\varphi^{2}/2\omega_\text{spp}^{2}T^{2}]$, 
with $ \varphi = k_{\text{spp}}x - \omega_\text{spp} t$. 
There are two counter-propagating SPP wave-packets, each containing a 
bright $\omega_b$ and and a dark $\omega_d$  frequency.
These terms are multiplied by a ``buildup'' and ``decay'' function $f(t) = 
\exp(-t/2\tau_m)\times(1-\text{erf}((\sigma_m^2-2\tau_m t)/(2\sqrt{2}\sigma_m\tau_m)))$, 
which is the convolution of a Gaussian excitation profile with exponential decay. 
Source duration and plasmon mode decay rate are denoted by $\sigma_m$ and $\tau_m$, respectively, for $m = b,d$. 
When $f(t)$ multiplies the plasmonic term, the respective $\tau_m$ parametrizes 
the lifetime, while the Gaussian half-width half-maximum in intensity 
$\xi_m = \sigma_m\sqrt{\ln 2}$ parametrizes the buildup time.

The remaining fit parameters are the amplitudes of the respective plasmon modes. 
The explicit form of the parametrization is given in the supplementary materials. 
The relevant free parameters in this model are the excitation buildup times $\xi_b,\xi_d$, 
the plasmon decay times $\tau_b,\tau_d$ and the plasmon frequencies $\omega_b,\omega_d$
for the bright and dark modes, respectively. 

Fitting to the simulated variance, we find plasmon frequencies are 
$\hbar\omega_b = 1.65$ eV and $\hbar\omega_d = 1.62$ eV, consistent with the 
plasmonic band gap of 14 nm given in Ref.~\cite{ropers2007njp}.
Results for the buildup- and life-times are reported in Table \ref{tab:fit-res}.
Because of spatial integration, the plasmon pulse extension $T$ has little influence on the  
variance. The values in the table were obtained with $T=15$ fs (FWHM). 
A conservative lower bound of $T$ is given by the diameter of the NIR spot size, an upper 
bound by that size plus plasmon propagation during excitation. 
\change{Variation in the range of $T=10$ and $20$ fs has only a small effect on 
buildup and decay times. Due to the superposition of bright mode decay with dark mode buildup, variation is largest for these
parameters with about 0.7 fs.}
For any given value of $T$ in this interval, the buildup and decay extracted from the FDTD surface field and from the spectrogram variance are in good agreement.
\begin{table}[t]
  \caption{\label{tab:fit-res}  
    Carrier frequency $\omega_m$, buildup time $\xi_m$ and lifetime $\tau_m$ 
    resulting from fits of the theoretical 
    model to the numerically simulated data.
    The cases isotropic emission with perpendicular detection (``filtered''), 
    unidirectional emission, as well as values extracted directly from the FDTD 
    calculation are shown. (Times in fs. Frequencies in eV)}
\begin{tabular}{cccccc}
  \hline
  \hline
  & $\quad$ & Filtered Isotropic & Unidirectional & FDTD\\
  \hline
  $\xi_b$    & $\quad$ & 2.07  & 2.06  & 2.01  \\
  $\tau_b$ & $\quad$ & 3.0   & 3.1   & 2.96  \\
  $\xi_d$    & $\quad$ & 6.6   & 6.2   & 5.3   \\
  $\tau_d$ & $\quad$ & 32.5  & 33.3  & 34.6  \\
  $\omega_b$ & $\quad$ & 1.61  & 1.62  & 1.62  \\
  $\omega_d$ & $\quad$ & 1.65  & 1.65  & 1.65  \\
  \hline
  \hline
\end{tabular}
\end{table}

A comparison of the two spectrograms in Figure \ref{fig:pho-spe-iso}
of the NIR vs. the plasmonic field allows the evaluation of the field enhancement, 
which is in the present case $\sim$ 1. 
>From the spectrogram at grazing direction, we get a NIR pulse 
duration of $\Delta t_\text{fwhm} = 4.5$ fs, in good agreement with the 
$4.6$ fs from the FDTD code. Such a measurement provides an independent
\emph{in situ} diagnosis of the field distortions 
of the NIR field caused by the interaction with the grating.


In conclusion, we have shown how to obtain, with existing experimental instrumentation, 
direct, time-resolved images of the SPP surface field. Time resolution is determined by 
controlling the relative pulse delay. This allows the extraction of basic parameters such 
as SPP buildup and life times. Attosecond resolution, in our example, provides for
the distinction of bright and dark mode oscillations. 
The same setup also provides {\em in situ} diagnostics of the NIR pulse.

Once spatially separated XUV attosecond and NIR pulses become available,
one may resolve in space and time also other surface phenomena: by letting the NIR 
field excite a surface mode in some region, one can image  
SPP propagation along complex plasmonic wave\-guides or plasmonic switches by 
simply pointing the attosecond XUV pulse on the region of interest.

We are grateful to C. Ropers for useful discussions.
We acknowledge support by the DFG, by the excellence cluster ``Munich
Center for Advanced Photonics (MAP)'', by the Austrian Science
Foundation project ViCoM (F41), by the Landesexzellenzcluster
"Frontiers in Quantum Photon Science" and the Joachim Herz Stiftung.

\bibliographystyle{apsrev4-1.bst}

\bibliography{library.bib}

\end{document}


\title{Supplementary Material}

\author{Mattia Lupetti$^a$, Julia Hengster$^b$, Thorsten Uphues$^b$, and Armin
  Scrinzi$^a$} \email{mattia.lupetti@physik.uni-muenchen.de}
\email{armin.scrinzi@lmu.de} \affiliation{$^a$Physics Department,
  Ludwig Maximilians Universit\"at, D-80333 Munich, Germany}
\affiliation{$^b$Center for Free-Electron Laser Science, Universit\"at
  Hamburg, D-22761 Hamburg, Germany}

\maketitle

\section{Parameterization of the plasmon}
The variance of the photoscopic spectrogram according to our theory is given by:
\begin{equation}\label{eq:var}
  S(\tau) = \Delta p^2 + \frac{1}{\pi}\int_{-\infty}^{\infty} dx_i\,
  g_\text{x}(x_i)\,\mathbf{a}_y^2(x_i, \tau).
\end{equation}
In order to compute the spatial integral in 
Eq.~(\ref{eq:var}), we consider a case where the bright and dark modes
can both be excited at frequencies  $\omega_b$ and $\omega_d$, respectively. ``Bright'' and ``dark''
 refer to the coupling properties of the modes: the bright mode 
 couples efficiently with incident radiation, the dark poorly.
The two frequencies are well separated. 
Each mode consists of two counter-propagating plasmons. In addition, we admit 
a term describing the ringing of localized modes excited in the focus of the NIR pulse
(see, e.g., [1]).
The contributions in each mode $m=b,d$ are
\begin{equation}
 P_m^{(\pm)}=e^{i\varphi_{\pm m}}e^{-\frac{\varphi_{\pm m}^{2}}{2\omega_m^{2}T_m^{2}}}
\end{equation}
with  the phase of the propagating plasmon 
\begin{equation}
\varphi_{\pm m} = \pm k_m x - \omega (t-t_m)
\end{equation} 
and 
\begin{equation}
P_m^{(0)}= \cos(\omega_m (t-t_m))\,e^{-x^2/2w_\text{nir}^2},
\end{equation}
for the localized excitation.

Buildup and decay are assumed to obey a simple rate equation where a Gaussian-shaped buildup of width 
$\sigma$ is depleted
by decay at a constant rate $\tau$:
\begin{equation}
\dot{f}(t)= e^{-\frac{t^2}{2\sigma^2}} - \frac{1}{2\tau}f(t).
\end{equation}
The resulting time-distribution is

\begin{align}\label{eq:conv}
f(t,\sigma,\tau) &= \int_0^t \, e^{-\frac{t'^2}{2\sigma^2}}e^{-\frac{t-t'}{2\tau}}\,dt'\nonumber\\
&= e^{\frac{\sigma^2-4 \tau  t}{8 \tau ^2}} \left[1-\text{erf}\left(\frac{\sigma^2-2 \tau  t}{2 \sqrt{2} \tau  \sigma}\right)\right].
\end{align}
With this, the complete field is parametrized by
\begin{align}\label{eq:tot-fie}\nonumber
\mathsf{a}_y(x,t)=&\sum_{m=b,d}f(t\!-\!t_m,\sigma_m,\tau_m)\nonumber\\
&\times\left\{ \text{a}_m\left[ P_m^{(+)}\!-\! P_m^{(-)} \right]+\text{c}_mP_m^{(0)}\right\}.
\end{align}
In practice, we find that the bright mode decays so fast that its propagation can be neglected 
in the spectrogram variance and we set a$_b\equiv0$.

We define the buildup time of each mode $\xi_m = \sigma_m \sqrt{\ln(2)}$ as the half-width half-maximum of the 
Gaussian function in Eq. (\ref{eq:conv}), which allows for a direct 
comparison with the NIR pulse FWHM duration.

The dark mode plasmon duration $T$ has a measurable effect only during generation,
when counter-propagating SPPs have not separated yet and form a standing wave.
As this process is superposed by the bright mode, it cannot be reliably retrieved from the fit. 
However, $T$ is only weakly correlated with the dynamical parameters $\xi_m,\tau_m$ and 
$\omega_m$. Table~\ref{tab:fit-res-fdtd} shows the dynamical parameters for variations of $T$ 
over $[10, 20]$ fs (FWHM).

\begin{table}[h]
  \caption{\label{tab:fit-res-fdtd}  
Buildup-, life-time, and frequency of the bright and dark modes as obtained by fitting 
Eq.~(\ref{eq:var}) with the parameterization (\ref{eq:tot-fie}), 
for a range of plasmon durations $T_m$. Times in fs, frequencies in eV, $T_{\text{FWHM}}=2\sqrt{\ln2}T$. }
\begin{tabular}{ccccccccc}
  \hline
  \hline
  $T $ & $6$ & $7$ & $8$ & $9$ & $10$ & $11$ & $12$ & Var \\
  $T_\text{FWHM}$ & $ 9.99$ & $11.66$ & $13.32$ & $14.99$ & $16.65$ & $18.32$ & $19.98$ & \\
  \hline
  $\xi_b$     & $1.933$ & $1.964$ & $1.987$ & $2.004$ & $2.016$ & $2.025$ & $2.031$ & $5$ \%   \\
  $\tau_b$  & $3.285$ & $3.137$ & $3.031$ & $2.964$ & $2.924$ & $2.903$ & $2.896$ & $13$ \%  \\
  $\xi_d$     & $5.941$ & $5.649$ & $5.430$ & $5.286$ & $5.202$ & $5.160$ & $5.149$ & $15$ \%  \\
  $\tau_d$  & $34.18$ & $34.41$ & $34.57$ & $34.63$ & $34.62$ & $34.57$ & $34.47$ & $< 1$ \% \\
  $\omega_b$  & $1.613$ & $1.615$ & $1.616$ & $1.617$ & $1.617$ & $1.616$ & $1.615$ & $< 1$ \% \\
  $\omega_d$  & $1.645$ & $1.645$ & $1.645$ & $1.645$ & $1.645$ & $1.645$ & $1.645$ & $< 1$ \% \\
  \hline
  \hline
\end{tabular}
\end{table}

\rule{0cm}{0cm}\\
\vspace*{\fill}\\
{[1]} F.C. Garcia-Vidal {\it et al.}, J. Lightwave Technology {\bf 17}, 2191 (1999).